\documentclass[fleqn,usenatbib]{mnras}

\usepackage{newtxtext,newtxmath}

\usepackage[T1]{fontenc}

\DeclareRobustCommand{\VAN}[3]{#2}
\let\VANthebibliography\thebibliography
\def\thebibliography{\DeclareRobustCommand{\VAN}[3]{##3}\VANthebibliography}

\usepackage{graphicx}	
\usepackage{amsmath}	

\title[PUMPS1]{Pushchino multibeam pulsar search: I. Targeted search of weak pulsars}

\author[S. A. Tyul'bashev et al.]{
Sergei A. Tyul'bashev,$^{1}$\thanks{E-mail: serg@prao.ru (SAT)}
Marina A. Kitaeva,$^{1}$
and Gayane E. Tyul'basheva$^{2}$
\\
$^{1}$ P.N. Lebedev Physical Institute of the Russian Academy of Sciences, Astro Space Center, Pushchino Radio Astronomy Observatory,\\
Radiotelescopnaya 1a, Moscow reg., Pushchino, 142290, Russia \\
$^{2}$ Institute of Mathematical Problems of Biology, brunch of Keldysh Institute of Applied Mathematics,\\  
Vitkevich 1, Moscow reg., Pushchino, 142290, Russia\\
}

\date{Accepted XXX. Received YYY; in original form ZZZ}

\pubyear{2015}

\begin{document}
\label{firstpage}
\pagerange{\pageref{firstpage}--\pageref{lastpage}}
\maketitle

\begin{abstract}
The search for pulsars in a sample of pulsar candidates found based on a multi-year survey conducted with low (6 channels; sampling 0.1s) time-frequency resolution on declinations $-9^o < \delta < +42^o$ was carried out. A Large Phased Array (LPA) transit telescope operating at 111 MHz in the 2.5 MHz band was used. Search, analysis and evidence of pulsar detection were carried out using a visualization program of summed up power spectra obtained from the survey data with high 32 channels; sampling 12.5ms) time-frequency resolution. 11 new pulsars with periods $P_0=0.41-3.75$~s and dispersion measure $DM = 15-154$~pc/cm$^3$ have been discovered. In total, in the survey with a low time-frequency resolution for the period 2016-2021 in a blind search, 208 pulsars were found, of them 42 new and 166 known pulsars. It is shown that in the search on the data with high time-frequency resolution accumulated over a time interval of 7 years, pulsars with a flux density of $0.1-0.2$~mJy at the frequency of 111~MHz can be detected. When searching for pulsars with regular (periodic) emission at declinations $+21^o< \delta < +42^o$, all pulsars located outside the galactic plane, having $P_0 \ge 0.5$~s, $DM \le 100$ pc/cm$^3$  and the flux density $S \ge 0.5$~mJy, can be detected. 

\end{abstract}




\section{Introduction}

To date, there are almost 3300 pulsars in ATNF catalogue\footnote{https://www.atnf.csiro.au/research/pulsar/psrcat/}  (\citeauthor{Manchester2005}, \citeyear{Manchester2005}). However, these pulsars are less then 10\% of expected Galactic population neutron stars visible as pulsars (\citeauthor{Keane2008}, \citeyear{Keane2008}).The discovery of an object with a period of 18.8 minutes (\citeauthor{Hurley-Walker2022}, \citeyear{Hurley-Walker2022}) shows the poverty of our knowledge about pulsars. In addition to new types of pulsars, new surveys allow us to study pulsar populations, their spatial distribution, the interstellar medium, and much more.

Over the past 10 years, there has been a surge in the pulsars search activity. Surveys have been conducted at both high and low frequencies (see, for example, \citeauthor{Keith2010} (\citeyear{Keith2010}), \citeauthor{Deneva2013} (\citeyear{Deneva2013}), \citeauthor{Boyles2013} (\citeyear{Boyles2013}), \citeauthor{Barr2013} (\citeyear{Barr2013}), \citeauthor{Stovall2014} (\citeyear{Stovall2014}), \citeauthor{Tyulbashev2016}, (\citeyear{Tyulbashev2016}), \citeauthor{Tyulbashev2018a} (\citeyear{Tyulbashev2018a}), \citeauthor{Sanidas2019} (\citeyear{Sanidas2019}), \citeauthor{Good2021} (\citeyear{Good2021}), \citeauthor{Amiri2021} (\citeyear{Amiri2021}), \citeauthor{Han2021} (\citeyear{Han2021})). This surge is associated with the emergence of new tools with a large effective area (LOFAR, CHIME, FAST), new recorders that allow digitizing wide frequency bands with the subsequent creation of a large number of frequency channels, as well as using new methods of interference suppression. A large effective area, wide frequency bands, and an increase in accumulation time lead to an increase in sensitivity in surveys, and as a result, to the detection of new pulsars.

The search for pulsars in the surveys is carried out using new or known, but not previously used processing methods (e.g., \citeauthor{Eatough2010} (\citeyear{Eatough2010}), \citeauthor{Cameron2017} (\citeyear{Cameron2017}), \citeauthor{Tyulbashev2017} (\citeyear{Tyulbashev2017}), \citeauthor{Cadelano2018} (\citeyear{Cadelano2018})). Archived data from previously conducted surveys are also being reprocessed. A simultaneous use of several search methods has become the standard in the search for pulsars. The classical search for harmonics in power spectra or summed power spectra obtained using the Fast Fourier Transform (FFT) is used, followed by non-coherent summation of harmonics. A search is performed using the Fast Folding Algorithm (FFA). Periodograms have not been actively used in the search for pulsars in surveys across all the sky before, since they require more computing resources than FFT. The advantage of FFA over FFT is that in FFA a coherent summation of all pulsar pulses is carried out. At the same time, higher sensitivity is realized as the outcome, especially for pulsars with long periods or high duty cycle (see handbook \citeauthor{Lorimer2004} (\citeyear{Lorimer2004}) and references in it). A search is also carried out for single dispersed pulses that can be associated with rotating radio transients (RRAT) or with fast radio bursts (FRB)  (\citeauthor{McLaughlin2006}, \citeyear{McLaughlin2006}, \citeauthor{Lorimer2007},  \citeyear{Lorimer2007}). 

After processing the initial data, a large number of candidates with attributes of pulsars appear. These candidates must either be confirmed by obtaining an average profile and a dispersion measure ($DM$) of pulsar, or rejected by showing that this candidate is a false detection. For confirmed pulsars, as a rule, additional observations are carried out, which make it possible to more accurately estimate the period ($P$), the period derivative ($\dot P$) and the coordinates. 

A sharp increase in the number of pulsar candidates observed in recent years was shown in publication \citeauthor{Lyon2016} (\citeyear{Lyon2016}). If in the earliest surveys one candidate per ten square degrees was found, then in the latest surveys processing programs generate from several thousand to almost one hundred thousand candidates per square degree. Such a large number of candidates cannot be checked visually, therefore, additional ways of screening out candidates and speeding up the search are being developed (\citeauthor{Eatough2010} (\citeyear{Eatough2010}, \citeauthor{Devine2016} (\citeyear{Devine2016}), \citeauthor{Bethapudi2018} (\citeyear{Bethapudi2018}), \citeauthor{Morello2019} (\citeyear{Morello2019})). 

The result of new observations, new processing methods and new methods of analysis of the processed data is pulsars discoveries. These pulsars are detected in areas of the sky, which have already been searched for pulsars several times. A striking example is a survey conducted on a 500-meter FAST mirror (\citeauthor{Han2021}, \citeyear{Han2021}). The effective area of FAST is about 2 times the area of the telescope in Arecibo and 20 times the area of the 64-meter telescope in Parks. At a partially processed area near the Galactic plane, 201 new pulsars were detected. 

A natural question arises about the number of pulsars in the sky available for observation. Such evaluations were made of course, (see, for example, \citeauthor{Keane2008} (\citeyear{Keane2008})), however, when estimating this amount of pulsars observed in the radio range, the luminosity function obtained from incomplete samples is used. Moreover, when conducting new surveys, it turns out that previously detected pulsars are often not detected in a blind search (see, for example, \citeauthor{Manchester1996} (\citeyear{Manchester1996})), which raises the question of the declared sensitivity of surveys. Therefore, the question of the number of pulsars in the radio range in the sky available for observation remains open. 

In 2014, monitoring (round-the-clock) observations under the “Space Weather” program were started at the radio telescope Large Phased Array (LPA) of the Lebedev Physical Institute (LPI) of the Russian Academy of Sciences (\citeauthor{Shishov2016}, \citeyear{Shishov2016}). In 2015, data obtained with low time-frequency resolution was used to search for pulsars.  24 days of observations were processed and 7 new pulsars (\citeauthor{Tyulbashev2016}, \citeyear{Tyulbashev2016}) were detected in the area $+21^o < \delta < +42^o$. Since 2014, observations are carried out around the clock (with small interruptions due to accidents and maintenance on the antenna).

To increase the sensitivity when searching for pulsars, we started using the summation of power spectra, which allowed us to detect 17 pulsars in the area $-9^o < \delta < +42^o$ (\citeauthor{Tyulbashev2017}, \citeyear{Tyulbashev2017}). Subsequent improvement of the processing technique made it possible to detect 5 more pulsars in the summed-up spectra at the same area (\citeauthor{Tyulbashev2020}, \citeyear{Tyulbashev2020}). At the same time, as noted in this publication, we were unable to verify some of candidates, since we could not find a single day for them in the original (raw) data to confirm them. 

We were unable to verify 135 candidates from our previous papers. These candidates may be known pulsars detected in the side lobes of the antenna, periodic interference, or new pulsars. Here, data with high time-frequency resolution is used to search of pulsars. These data make it possible to realize a higher sensitivity, since pulsar pulses are less smoothed inside the frequency channels, and the sampling time is less than the pulse duration.

This publication is the final one for searching of pulsars in the data with low time-frequency resolution. A new way of visualizing the processed data used in the weak pulsar search program is discussed in it, and pulsars detected using this program are shown. We also consider the sensitivity of the LPA when searching for pulsars in data with high time-frequency resolution.

The paper is organised as follows: the observation in Sec.2; data processing and analysis in Sec.3; results of pulsar search in Sec.4; questions about sensitivity of survey in Sec.5; some discussion in Sec.6; conclusions in Sec.7.

\section{Observations}

The main purpose of Pushchino Multibeam Pulsar Search (PUMPS) – is a search for pulsars and fast transients (RRAT and FRB) in the data obtained on the meridian-type radio telescope LPA LPI, as well as the study of open and known sources using the same data. The antenna consists of 16384 wave dipoles. Using the antenna field and multiplication of signals at the input of the cable system we managed to create in 2012 four independent beam systems based on Butler matrices  (\citeauthor{Shishov2016}, \citeyear{Shishov2016}). Two of them are used in the survey for the search and study of pulsars and transients. One of these beam systems, LPA3 , was created for observations under the “Space Weather” program. It consists of 128 beams fixed in the direction, and overlaps declinations $-9^o <\delta < +55^o$ (see Fig.~1 in \citeauthor{Shishov2016} (\citeyear{Shishov2016})). Until 2021, 96 beams out of 128 were connected to two recorders overlapping declinations from $-9^o$ to $+21^o$ and from $+21^o$ to $+42^o$, and since January 2021, another recorder has been added to them, handling 24 beams overlapping declinations from $+42^o$ to $+52^o$. The beam size for the LPA antenna is approximately $0.5^o \times 1^o$, by declination and right ascension. The time taken for a beam to transit a point in sky at the half power is approximately 3.5 minutes at declination $\delta = 0^o$.  The central frequency of observations is 110.3~MHz, the observation band is 2.5~MHz. The coordinates of the 120 beams used are fixed and overlap declinations from $-9^o$ to $+52^o$.

At LPA3, a daily (monitoring) survey of the sky are conducted, used for a number of scientific tasks, including the search for pulsars and transients (\citeauthor{Shishov2016}, \citeyear{Shishov2016}, \citeauthor{Tyulbashev2016}, \citeyear{Tyulbashev2016}, \citeauthor{Tyulbashev2018a}, \citeyear{Tyulbashev2018a}). For pulsar observations, the typical sensitivity in one observation session is estimated as 6-8~mJy for observations outside the Galactic plane and 15-20~mJy for observations in the Galactic plane (\citeauthor{Tyulbashev2016}, \citeyear{Tyulbashev2016}). 

To conduct a survey at declinations above $+55^o$, the LPA1 radio telescope is used. On this telescope, neighboring beams overlap at the level of 0.81. Therefore, for a source that falls exactly between the beams of the LPA LPI, 0.81 of the total energy is observed in neighboring beams. The possible declinations of LPA1 are from $-16^o$ to $+87^o$. In practice, negative declinations are rarely used, because the lower the declination, the greater the level of interference from the nearest major cities of Moscow and Tula, located approximately 100 km away to the north and south of the city of Pushchino. In addition, the lower the declination, the smaller the effective area of the antenna. At the declination $\delta = -5^o$, the effective area of the LPA drops by half. A test observations by LPA1 began in April 2019. The Program Committee allocates approximately 10 days each month for the monitoring survey. The LPA1 recorder allows to record observations in eight beams aligned in the meridian plane, to set different digitization times for one point (sampling) and a different number of frequency channels.

PUMPS includes three big search tasks. The first one is the search for seconds duration pulsars at the sensitivity better than 1~mJy at the frequency of 111~MHz. The second one is the search for rotating radio transients (RRAT). The third is the search for fast radio bursts (FRB). In this paper, the first task is considered in the application to the observations on LPA3.

Raw data from LPA3 are recorded around the clock in parallel in two modes. In the 6-channel mode, data with low time-frequency resolution has a volume of 1~TB per year. In this mode, the sampling time is $\Delta t = 0.1$~s, and the channel width is $\Delta \nu = 430$~kHz. The data of this mode was originally intended for the “Space Weather” program, but we also used it to search for pulsars (\citeauthor{Tyulbashev2016}, \citeyear{Tyulbashev2016}, \citeauthor{Tyulbashev2017}, \citeyear{Tyulbashev2017}, \citeauthor{Tyulbashev2020}, \citeyear{Tyulbashev2020})).  
The second mode is 32-channel one. In this mode, data has a volume of 35~TB per year. This is data with high time-frequency resolution, they are recorded with $\Delta t = 12.5$~ms and $\Delta \nu =78$~kHz. Such a sampling time and such a channel width, of course, are not sufficient for a full-fledged search for seconds duration pulsars, but still data with high time-frequency resolution are much better than data with low time-frequency resolution. Here and further in the text, we will use the conditional names of data with low and high time-frequency resolution as "ltfr data" and "htfr data".

Observations on LPA3 were not originally intended to search for pulsars, since the recorder was made to work on the "Space Weather" task. This recorder has problems with the accuracy of timestamps. The recording goes in hourly segments, and at the end of the hour recording, the next hour segment is immediately started. The beginning of the hour recording is controlled by the time service, which also provides regular pulsar observations. Inside the time interval, the time is tracked by a quartz oscillator. For htfr data, the accuracy of the time measurement running away or lagging time is approximately $\pm 25$~ms, for ltfr data it is $\pm 100$~ms. This accuracy of the timestamps was ensured when checking the recorders in the laboratory, but was not verified in real observations.

\section{Data processing and analysis}

Summed power spectra for all the days of observations are used the increase in sensitivity (\citeauthor{Tyulbashev2017}, \citeyear{Tyulbashev2017}, \citeauthor{Tyulbashev2020}, \citeyear{Tyulbashev2020}). If the recorded background noise at a given point in the sky does not change from one observation session to another over the entire observation interval and there are no changes in the pulsar flux density over time, then the sensitivity should grow as the square root of the number of observation sessions. In reality, there are changes in the recorded noise of the radio telescope depending on a weather, a physical condition of the antenna, the antenna-feeder path, a total amount of interferences on a given day, and other factors, i.e. there is a change in sensitivity. As a rule, the experimental sensitivity assessment is about 20-30\% lower than the theoretical one (\citeauthor{Tyulbashev2020}, \citeyear{Tyulbashev2020}).

Previously, the search for pulsar candidates was carried out in the summed-up power spectra obtained from ltfr data, and the detected candidates were checked using htfr data. The length of the ltfr data record being processed is 2048 points (204.8~s), which is approximately equal to the dimension of the LPA beam at half power. 

Objects for which harmonics with signal to noise ratio $(S/N)>5$ were detected in 4-year averaged power spectra were considered as candidates for pulsars. For these candidates, we tried to find at least one session in the htfr data in which it would be possible to obtain an average profile, estimate the $DM$ and clarify the period ($P_0$) of the pulsar. 

Obviously, there will come a time when searching for pulsars by harmonics in the summed-up power spectra, it will be possible to detect new pulsars by their accumulated signals, but they cannot be seen in single sessions due to the insufficient sensitivity of a radio telescope. For such weak pulsars, processing confirming the characteristics ($P_0$ and $DM$) of new pulsars without getting their averaged profiles is necessary.

In previous papers, when processing ltfr data, we used the BSA-Analytics program\footnote{https://bsa-analytics.prao.ru/en/downloads.php} to obtain its own power spectrum for each frequency channel, then summed-up the power spectra across all channels. Such processing was done for each day of observations and for all available directions in the sky. The power spectra obtained for different days were summed up. With this method of accumulating power spectra, a minimum number of output files is formed, equal to the number of directions available in the sky. Then we looked for harmonics in the summed-up spectra followed by visual verification. Information about the $DM$ of the pulsar candidate is lost with this approach. This method made it possible to process as quickly as possible several years of daily observations of the area occupying in the sky 17000 sq.degrees (see additional details in publications \citeauthor{Tyulbashev2017} (\citeyear{Tyulbashev2017}), \citeauthor{Tyulbashev2020} (\citeyear{Tyulbashev2020})). 

In total, we have discovered 30 new pulsars in the ltfr data (\citeauthor{Tyulbashev2016}, \citeyear{Tyulbashev2016}, \citeauthor{Tyulbashev2017}, \citeyear{Tyulbashev2017}, \citeauthor{Tyulbashev2020}, \citeyear{Tyulbashev2020}). At the same time, there are more than a hundred candidates left in our lists, which we could neither identify with the pulsars known for 2019, nor show that the candidate under study is a new pulsar. It was necessary to develop a method by which it would be possible to confirm or dispose of the detection of a new pulsar for each of these candidates.

For this purpose, a new processing program was developed, which in the raw data processing module for each pulsar candidate generates power spectra sorting out different $DM$, averaging of raw data and offsets of the pulsar coordinates. Then the power spectra obtained for each day are summed-up over several years, separately for each set of parameters being sorted out. 

That is, first, for a given direction (coordinates for declination and right ascension) and a given day (date), power spectra are calculated, taking into account $DM$ compensation, sorted out within range $0 \le DM \le 1000$ pc/cm$^3$. We use a DM step equal to one at small DM and increase it in proportion to the broadening of the pulse due to smearing in the frequency channals.

To evaluate good (without interference) of the power spectra and to remove bad (with interference) spectra, we used the spectra obtained on $DM=0$~pc/cm$^{-3}$. The process of obtaining high-quality spectra is shown in detail in the work \citeauthor{Tyulbashev2020} (\citeyear{Tyulbashev2020}). We get each such power spectrum first without averaging, and then with a sorting out averaging of raw data in increments of 2 (1, 2, 4, 8, 16, 32 points). The averaging increment 1 - corresponds to the assumed width of the average pulsar profile 12.5 ms, the increment 32 - corresponds to the width of the average pulsar profile $12.5 \times 32 = 400$~ms. In addition, for each pulsar candidate, we get all these spectra not only for its assumed central coordinate, but also shifting from it by 1, 2, 3 minutes in both directions. As a result, when searching for pulsars for a given direction in the sky, processing one session, instead of one power spectrum in the old version of ltfr data processing, we get more than six thousand spectra in the new version of htfr data processing. 

Processing the data for the following days, we sum up the power spectra with the same set of positions for different days, and get the summed-up spectra for monthly and annual intervals, as well as for the full observation interval. In this work, the verification of pulsar candidates was carried out using power spectra summed-up for the htfr data interval available to us, i.e. for 5.5 years (August, 2014 – December, 2019).

The obtained power spectra are fed into the visualization module. It allows to view the summed-up power spectra for each pulsar candidate, check diagrams of the dependences $P_0/DM$, $(S/N)/DM$, $(S/N)/W_e$,  $(S/N)/dR$ ($dR$ – is a shift of the expected pulsar coordinate inside the radiation pattern), and based on the results of their analysis, we can conclude whether this candidate is a true pulsar. 

\begin{figure*}
	\includegraphics[width=\textwidth]{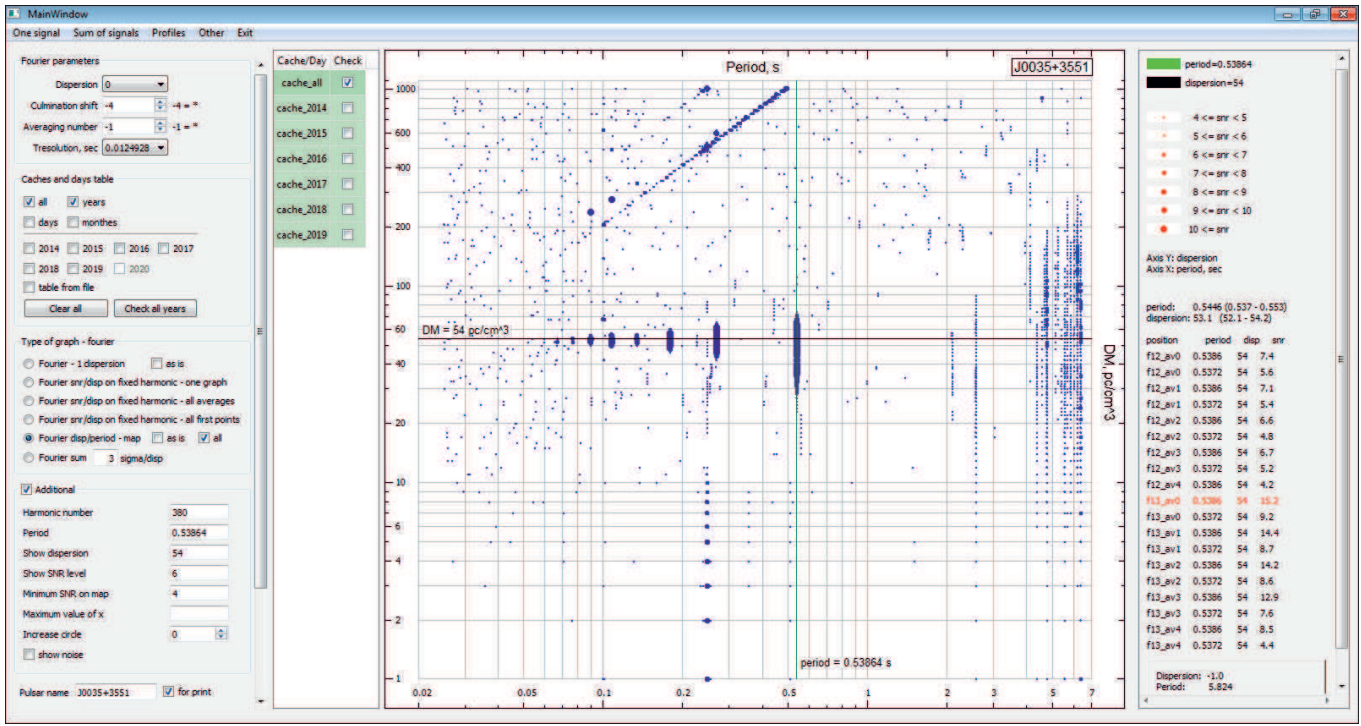}
    \caption{General view of the window of the power spectrum visualization module.}
    \label{fig:fig1}
\end{figure*}

In order to make it possible to capture and evaluate the pulsar candidate's picture, consisting of thousands of its power spectra, as simply and quickly as possible, the visualization program makes a map. On it, harmonics are shown in the summed-up power spectra by circles of different sizes – only those with $S/N >4$. The higher the $S/N$, the larger is the size of a circle. Along the vertical axis of the map, the $DM$ is shown with which the summing up was made for the channels of raw data, and along the horizontal axis, the period $P_0$  corresponding to the harmonic number in the summed-up power spectrum is shown (see Fig.~\ref{fig:fig1}). Harmonics of power spectra constructed from source data with different parameters may fall into the same point on the map and overlap each other, but information about them is preserved on the map.

If the object under study is a pulsar, then we will see a vertical strip on the map, thickening towards the center. The X-coordinate of this strip will be equal to the pulsar period, and the Y-coordinate can be used to estimate the $DM$. The maximum $S/N$ of a harmonic should be observed in the summed-up power spectrum obtained by summing up the frequency channels of raw data, taking into account the true $DM$ of the pulsar. On neighboring dispersions, the pulsar will also be observed, but with a lower $S/N$. Therefore, the center of the strip with the maximum thickness (i.e. with maximum $S/N$) will be at the Y-coordinate, equal to the pulsar dispersion, and up and down from the center the strip will thin out (because $S/N$ reduces). The height of the strip should depend on the width of the average profile. The narrower the average profile, the shorter the strip. If several harmonics are visible in the summed-up power spectrum, then several parallel strips will appear on the map.

The map is interactive. When you click on a point on the map, information about this point appears in the right part of the window, it is a list of parameters of the summed-up power spectra that formed a circle at this point. The list includes a code of a power spectrum, its dispersion, a period corresponding to the harmonic that is shown on the map, and the $S/N$ of this harmonic. By clicking on the line, you can go to view its power spectrum, and then to other diagrams. 

Thus, having seen on the map an object similar to a pulsar, you can click on its center, then select the power spectrum with the best $S/N$ in the right part, and immediately proceed to viewing it. And then look at all its characteristics: a diagram of $S/N$ dependence of the selected harmonic on the dispersion, a change of this diagram depending on the averaging of raw data, and also on a time shift by $\pm$ 1,2,3 minutes, on assumed coordinate of right ascension. In this paper, we tested pulsar candidates found in a survey conducted earlier, but the search methodology developed on this sample will be used in our planned blind search across the sky.

The power spectrum is determined on the array with the length $204.8$~s $=12.5$~ms~$\times 16384$~points, which is approximately equal to the time of passage of the source through half of the radiation pattern of the LPA LPI. During the day (24 hours), there are $(24\times 3600)/204.8=422$ such segments. The monitoring observations are carried out in 96 beams of LPA3, covering the sky along the meridian, therefore, the whole sky in the range of declinations $-9^o<\delta < +42^o$ can be represented as $422 \times 96 \simeq 40500$ of beam directions, covering the area of 17000 sq. degrees. One of the directions is the area of 17000/40500=0.4 sq. degrees. Thus, the entire sky available for observation can be represented as 40500 pixels, each of which corresponds to a certain position of the LPA LPI beam. In the planned search, the visualization of the summed-up power spectra for each of these pixels will be carried out. 

Summed-up periodograms obtained using the Fast Folding Algorithm (FFA) are also used for search for pulsars. 
We used summed FFA for pulsar candidates that were not detected on FFT maps. The logic of working with FFA in the program is similar to the logic of working with FFT. Using FFA, we create maps similar to the maps obtained from the summed-up power spectra, we get the $S/N$ for harmonic dependencies vs. $DM$ and all other dependencies similar to the search using FFT. No candidates were confirmed through FFA. The part of the FFA program responsible for periodograms\footnote{https://github.com/v-morello/riptide} was taken from the paper \citeauthor{Morello2020} (\citeyear{Morello2020}). Some additional details can be found in the paper \citeauthor{Tyulbashev2021} (\citeyear{Tyulbashev2021}).

An example of the analysis of PSR J1745+12 using a new visualization program is given in APPENDIX A.

\section{Results}

The search for pulsars using summed-up power spectra from ltfr data was considered in the papers \citeauthor{Tyulbashev2017} (\citeyear{Tyulbashev2017}), \citeauthor{Tyulbashev2020} (\citeyear{Tyulbashev2020}), but, as was noted in these papers, there were candidates remaining for which it was not possible to obtain $DM$ estimates. In this paragraph, we discuss all the remaining pulsar candidates that have been tested using a new data processing and visualization program.

A total of 135 candidates were checked, in which harmonics were found having $S/N \ge 5$. Using the known coordinates of the right ascension and declination, summed-up power spectra were obtained from htfr data for these candidates on the interval 2014-2019 years. The summed-up spectra were fed into the visualization module of the program, in which their analysis was carried out using FFT maps. 

Out of the 135 candidates tested, 55 candidates in the data with htfr were not confirmed – it turned out that the harmonics observed were caused by interference 31 candidates were identified using $P/DM$ maps with one or two pulsars detected in the side lobes of the LPA LPI. Previously, we were unable to detect these pulsars in the side lobes of the LPA, since the second or subsequent harmonics of the pulsar were often visible in the power spectrum obtained from ltfr data. At the same time, the pulsar appeared in a side lobe, the coordinates of which could differ by tens of degrees from the coordinates of the pulsar that gave birth to this lobe. Without $DM$ estimates, it was impossible to select a candidate for possible identification from the ATNF catalog. The visualization program made it easy to identify these pulsars by their $DM$ and period.

For 34 candidates, known pulsars were detected on the maps, which were not detected in previous studies (and in some cases there were two such pulsars). Separately, we note a number of pulsars discovered by different authors after 2013, that is, after the start of the PUMPS survey, and not confirmed by us earlier due to the lack of a $DM$ estimate. It is pulsars: PSR J0358+4155, PSR J0612+37216 and PSR J2105+28, discovered on a 100-meter telescope Green Bank (\citeauthor{Stovall2014}, \citeyear{Stovall2014}, \citeauthor{Aloisi2019}, \citeyear{Aloisi2019}), as well as such pulsars as PSR J0035+35, PSR J0349+2340, PSR J0813+22, PSR J1707+35, PSR J1745+1252, PSR J1814+22, PSR J1849+2559, PSR J2006+22, PSR J2022+21, PSR J2123+36, discovered on LOFAR (\citeauthor{Sanidas2019}, \citeyear{Sanidas2019}). We confirm their discovery on the LPA LPI. These pulsars are not the subject of this publication, but some information about them is available on our website\footnote{https://bsa-analytics.prao.ru/en/pulsars/known.php}, where are the maps exported from the visualization program.

\begin{figure}
\begin{center}
	\includegraphics[width=0.8\columnwidth]{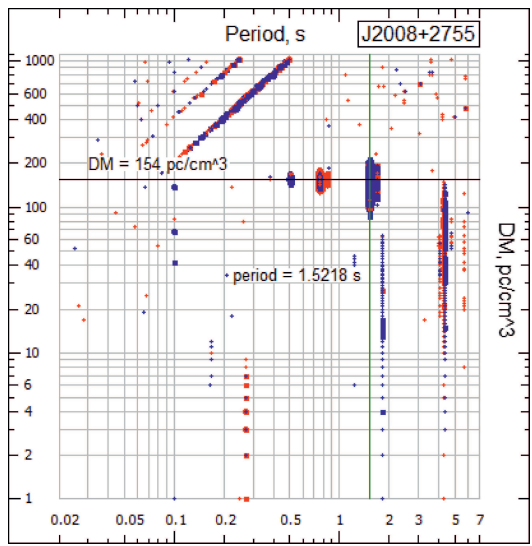}
    \caption{The map shows the pulsar PSR J2008+2755, which has the maximum $DM$ among all pulsars discovered in the survey. On the vertical axis is $DM$, on the horizontal axis is period. The scale of both axes is logarithmic.}
    \label{fig:fig5}
\end{center}
\end{figure}

In this sample, we found 11 new pulsars. Fig.~\ref{fig:fig5} shows the $P/DM$ map for the new pulsar PSR J2008+2755, which has the largest $DM$ of all the pulsars found. The pulsar is located in the Galactic plane (galactic latitude $b<-3^o$) where, as stated in the previous paragraph, the sensitivity of PUMPS is 2.25 times worse than the sensitivity outside the Galactic plane. The processing of observations for PSR J2008+2755 was done out for 865 days. The map drawn according to htfr data shows that for the first harmonic $S/N=18.1$, $DM=154$~pc/cm$^3$, $P_0=1.5218$~s. All signals with $S/N \ge 5$ are displayed on the map. For the pulsar on the line $DM = 154$~pc/cm$^3$, three vertical stripes and one point are visible - these are the first four harmonics in the power spectrum of this pulsar.


The 13 candidates found in the summed-up spectra obtained from ltfr data are absent in the spectra and periodograms summed-up from htfr data. It was not possible to find out the reasons why harmonics that were clearly visible when processing ltfr disappeared when processing htfr data.

The expected flux densities of new pulsars may be lower than the sensitivity level achieved in a single observation session. It is possible to estimate $S/N$ in a single session for the weakest pulsars detected at $S/N=6$ in the summed power spectra if the pulsar flux density does not change with time: $(S/N)_{one-session}=6/28=0.2$. Therefore, it is impossible to detect weak pulsars in single sessions.  However, pulsars can have both their own variability, as well as variability induced by interstellar plasma. Therefore, despite the fact that the average pulsar flux density may be too low to be detected by LPA3 in individual sessions, the apparent flux density may sometimes increase, increasing our chances of make an average profile.

\begin{figure}
\begin{center}
	\includegraphics[width=0.6\columnwidth]{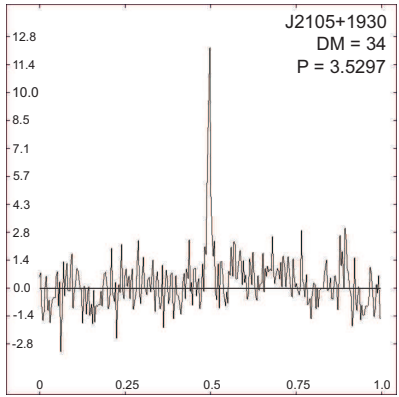}
    \caption{The averaged pulsar profile of PSR J2105+19 is generated by the processing program. Along the vertical axis, there is the amplitude in arbitrary units, along the horizontal axis, there is a phase.}
\end{center}
    \label{fig:fig6}
\end{figure}

The search for days when pulsars were strong enough to be detected in one observation session was carried out according to the following scheme: a) gain equalization was performed in 32-frequency channels of the data; b) a data segment of 16384 points was cut out (approximately 204.8 s) centered at the culmination point of the supposed pulsar; c) the power spectra were calculated and the days when $S/N>4$ in the harmonic corresponding to the pulsar period were determined; d) for the selected days, the periods and $DM$ (direct search) were sorted out near the values obtained in the search; e) since our computing resources do not allow us to check all the selected days, 30-50 days were checked for each candidate. The average profile shown in Fig.~\ref{fig:fig6} were obtained only for one pulsar (PSR J2105+19) out of 11 discovered ones. In Fig.~\ref{fig:fig6}, we present profile of PSR J2105+19 having maximum visible $S/N$.

Table~\ref{tab:tab2} shows estimates of parameters of discovered pulsars. The name of the pulsar is given in the first column, in columns 2-7 there are coordinates for right ascension and declination, period, $DM$, half-width of the pulse, estimate of the distance to the pulsar according to the YMW16 model\footnote{https://www.atnf.csiro.au/research/pulsar/ymw16/} (\citeauthor{Yao2017}, \citeyear{Yao2017}). An estimate of the distance to PSR J2253+12 shows that the YMW16 model needs improvement.

\begin{table*}
	\centering
	\caption{Characteristics of discovered pulsars.}
	\label{tab:example_table}
	\begin{tabular}{cccccccc} 
		\hline
name & $\alpha_{2000}$ & $\delta_{2000}$ & $P_0$(s) & $DM$~(pc/cm$^{-3}$) & $W_{0.5}$~(ms) &Dist~(kpc)\\
		\hline
J0109+11&01$^h$09$^m$45$^s$&11$^o$32$^\prime$&0.4327&17$\pm$2 & - &1.57\\
J0509+37&05 09 15&37 32&2.4961&30$\pm$3 & - & 0.92\\
J1844+21&18 44 45&21 51&0.5959&28$\pm$3 & - & 1.56\\
J184\bf{5}+21&18 45 00&21 51&3.7556&50$\pm$10& - & 3.38\\
J1917+17&19 17 30&17 24&0.4196&40$\pm$2 & - & 1.74\\
J1921+34&19 21 45&34 07&1.444 &85$\pm$5 & - & 7.29\\
J2008+27&20 08 15&27 55&1.5218&154$\pm$5& - & 7.52\\
J2029+34&20 29 45&34 37&1.8194&97$\pm$5 & - & 5.19\\
J2105+19&21 05 45&19 24&3.5297&35$\pm$3 &35 & 3.05\\
J2253+12&22 53 15&12 32&0.9793&37$\pm$5 & - & >25\\
J2333+20&23 33 45&20 18&2.2911&17$\pm$5 & - & 1.18\\		

		\hline
	\end{tabular}
	\label{tab:tab2}
\end{table*}

Table~\ref{tab:tab2} contains 11 new pulsars. When preparing the manuscript for submission to the journal, we found that in the GPPS survey (\citeauthor{Han2021}, \citeyear{Han2021}) the pulsar PSR J2008+2755 was detected having $P_0=1.51926$~s, $DM=155.1$~pc/cm$^3$. Apparently, it coincides with the PSR J2008+27 detected by us. It was also found that the pulsar PSR J2105+19 is close in terms of dispersion and coordinates to the RRAT J2105+19 (\citeauthor{Tyulbashev2018b}, \citeyear{Tyulbashev2018b}), whose period had not been determined. Apparently, we were able to detect the regular emission of this rotating transient.

\section{Sensitivity in PUMPS survey}

\subsection{Sensitivity for a low and high time-frequency resolution data}

Detecting new pulsars using htfr data gives indications that htfr data is better than ltfr data. Let's compare the sensitivity of surveys conducted with different time-frequency resolution. We consider both a theoretical sensitivity evaluation taking into account the width of the frequency channel, the scattering effect, the sampling used and changes in the sensitivity of the antenna array in different directions, as well as the test of the practical sensitivity of the survey based on known pulsars entering the observation area.

To calculate the sensitivity when searching for pulsars, we use the well known formula  (\citeauthor{Lorimer2004}, \citeyear{Lorimer2004}):\

\begin{equation}
    S_{min}=\frac{S/N \times T_{sys}}{G(n_p \times t \times \Delta \nu)^{1/2}} \times \sqrt{ \frac {W_e}{P-W_e}},
    \label{eq:1}
\end{equation}
where $S/N=6$ is minimal detected signal; $T_{sys}$ is a temperature of the system in the plane of the Galaxy and outside of it (for the evaluations $1800$K and $800$K); $G$ is a parameter related to the effective area of the antenna ($G=17$~K/Jy, \citeauthor{Tyulbashev2016} (\citeyear{Tyulbashev2016})); $n_p$ is a number of observed linear polarizations (in our case it is one polarization); $t$ is the duration of the observation session in seconds (for evaluations, we take it equal to $180$s); $\Delta \nu$ is full reception bandwidth in megahertz (2.5~MHz). $P$ and $W_e$  parameters are the expected period of the pulsar and the width of its average profile. For sensitivity evaluation, it is usually assumed that $W_e=0.1P$. However, the width of the average profile may become greater than $0.1P$ due to the smearing of the pulse in the frequency channel on large $DM$ and due to the scattering effect. Both effects will lead to a deterioration in sensitivity. For small $DM$, formula~\ref{eq:1}  will be valid, and for large ones, you need to take into account the additional broadening of the average profile:

\begin{equation}
W_e = (W_{0.1P}^2 + W_{\Delta t} ^2 + W_\tau ^2 + W_{DM} ^2 )^{1/2},
    \label{eq:2}
\end{equation}
where $W_{0.1P}$ is assumed width of the averaged profile; $W_{\Delta t}$ is raw data point sampling; $W_{DM}$ is pulse broadening due to smearing in the frequency channel associated with $DM$; $W_{\tau}$ is broadening of the averaged profile due to the scattering effect. The widths $W$ are defined in ms and expressed by the following formulas:

\begin{equation}
\begin{split}
W_{0.1P} = 0.1P; W_{\Delta t} = \Delta t; W_\tau = 60 \times (DM/100)^{2.2}; {}\\
 W_{DM} = 4.1488\times 10^6 \times (1/\nu_1 ^2 -1/\nu_2 ^2)\times DM {}, 
    \label{eq:3}
\end{split}
\end{equation}
where $\Delta t$ is sampling for one point, $\nu_1$ and $\nu_2$ are frequencies in MHz, determining the width of the frequency channel ($\Delta \nu = \nu_1 - \nu_2$). As a rule, in the surveys being conducted, the sampling of one point is in fractions of a millisecond, which is much less than the estimated pulse width of a seconds duration pulsar. In the data we use, the samplings of the point are 100 ms and 12.5 ms, the relation $0.1P \gg \Delta t$ it is not valid, and therefore, when evaluating the sensitivity, the sampling time must be taken into account. The expression for $W_\tau$  is taken from the paper \citeauthor{Kuzmin2007} (\citeyear{Kuzmin2007}). In this expression, scattering estimates were obtained from measurements at the frequencies 40, 60, 111 MHz. The resulting expression is very different from the one published in the paper \citeauthor{Bhat2004} (\citeyear{Bhat2004}) ($W_\tau \sim (DM)^{4.4}$), however for $DM<200$~pc/cm$^3$, the expressions approximately coincide. Low-frequency measurements of averaged profiles widenings carried out on LOFAR (\citeauthor{Geyer2017}, \citeyear{Geyer2017}), confirm that for pulsars with $DM<200$~pc/cm$^3$ , the scatterings are in agreement with the dependence given in the paper \citeauthor{Kuzmin2007} (\citeyear{Kuzmin2007}).

\begin{figure}
	\includegraphics[width=\columnwidth]{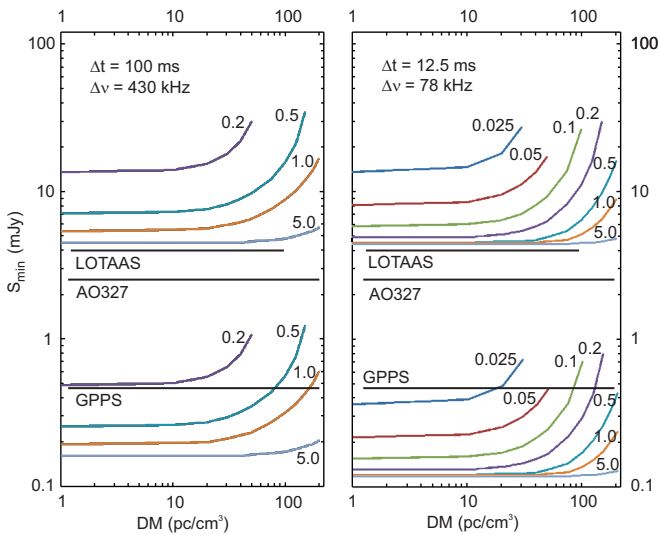}
    \caption{Dependence of sensitivity on the $DM$ for different periods in the PUMPS survey for ltfr data (left part) and htfr data (right part).  The upper part shows the sensitivities for one observation session, the lower part shows the sensitivity taking into account the summing up of the power spectra, in mJy. The colored curves are the sensitivities in the PUMPS survey for the periods indicated above the curves.}
    \label{fig:fig3}
\end{figure}

As \textbf{it} had been shown in the paper \citeauthor{Tyulbashev2020} (\citeyear{Tyulbashev2020}), when summing up the power spectra, the sensitivity grows slower than the square root of the number of summed-up power spectra. This is due to the fact that the sensitivity of the radio telescope may vary from one observation session to another due to weather conditions, breaks in dipole lines (especially in winter), untreated interference and other factors. In the paper \citeauthor{Tyulbashev2020} (\citeyear{Tyulbashev2020}), the dependences of the real increase in sensitivity for each pixel in the sky were developed. They show that the typical difference between experimentally determined sensitivity and theoretical one is in the range of 10-30\%. Here and below, for estimates, we will assume that the real sensitivity is 0.8 of the theoretical one, which in turn is defined as the square root of the number of summed-up power spectra or periodograms. During monitoring observations, part of the days are not processed. These days are associated with routine maintenance work at the LPA LPI, power grid accidents, thunderstorms, high solar activity, and other sources of interference. In general, we process at least 75\% of all days. Thus, it is possible to estimate the minimum increase in sensitivity in the summed-up power spectra and summed-up periodograms as:

\begin{equation}
S_{min-sum} \le \frac{S_{min}}{0.8 \times (0.75 \times N)^{1/2}} (mJy)
    \label{eq:4}
\end{equation}

where $N$ is the total number of days since mid-August, 2014, when the search for pulsars started. Fig.~\ref{fig:fig3} shows the sensitivity in observations at LPA3 for ltfr and htfr data for one observation session and taking into account incoherent summation of spectra or periodograms. The increase in sensitivity in our survey taking into account the summation of power spectra is $S_{min-sum-short} = 0.8 \times (0.75 \times 1595)^{1/2}=28$ times. Processing of htfr data is planned in 2022, by which time a number of observations will be more than 7 years long. Expected increase in sensitivity will be $S_{min-sum-long} = 0.8 \times (0.75 \times 2690)^{1/2}= 36$ times. We give the sensitivity outside the plane of the Galaxy. The sensitivity in the plane of the Galaxy is 2.25 times worse ($1800K/800K=2.25$), in order not to clutter up the figure, we do not show it.

Black horizontal lines are sensitivities $S_{min}$ of the best surveys (see next subsection) in the world at the moment on the search for pulsars, after translating their sensitivities  into 111~MHz. LOTAAS, AO327, GPPS are surveys done accordingly on LOFAR, Arecibo, FAST (\citeauthor{Deneva2013}, \citeyear{Deneva2013}, \citeauthor{Sanidas2019}, \citeyear{Sanidas2019}, \citeauthor{Han2021}, \citeyear{Han2021})).

A comparison of the left and right parts of Fig.~\ref{fig:fig3} shows that the shorter the period, the worse the sensitivity of the pulsar search survey. The sensitivities in the htfr and ltfr data survey are determined by the time of data sampling and the width of the frequency channel. For the ltfr data, the best sensitivity in the PUMPS survey is 14~mJy for pulsars with $P=0.2$~s and $DM \le 20$ pc/cm$^3$, at a higher values of  $DM$, the sensitivity deteriorates dramatically. For pulsars with $P>1$~s and with $DM=50$~pc/cm$^3$ the sensitivity is about the same - $S_{min}=6$~mJy, and is close to the theoretical one, equal to 4.4~mJy (\citeauthor{Tyulbashev2016}, \citeyear{Tyulbashev2016}), and for $DM >50$~pc/cm$^3$ the sensitivity begins to deteriorate sharply.

\begin{figure}
	\includegraphics[width=\columnwidth]{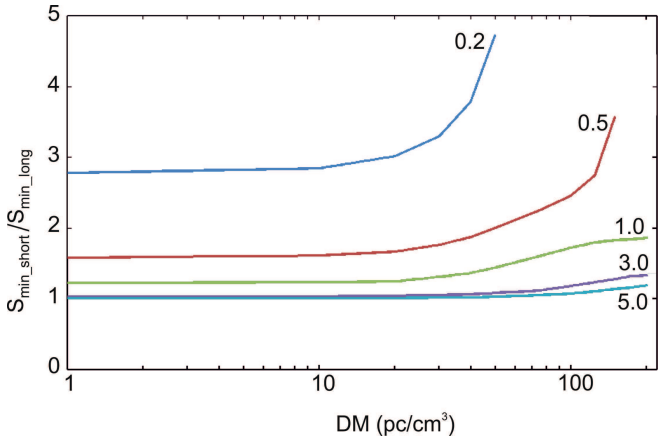}
    \caption{On the vertical axis there are numbers showing how many times the sensitivity in the survey on htfr data is better than the sensitivity in the survey on ltfr data ($S_{min-short}/S_{min-long}$). The horizontal axis shows $DM$. The number next to a curve shows a period in seconds.}
    \label{fig:fig4}
\end{figure}

For the htfr data, a decrease in the sampling time and frequency channel width leads to an improvement in sensitivity when searching for pulsars with small periods and large $DM$ compared to the sensitivity of the survey on the ltfr data. From the sensitivity (Fig.~\ref{fig:fig3}) for the ltfr data, it can be seen that when searching for pulsars with $0.2$~s $\le P \le 1$~s and $0 \le DM \le 200$~pc/cm$^3$ the broadening of the pulse associated with the large width of the frequency channel will lead to the fact that already at  $DM=100$~pc/cm$^3$, pulsars with $P_0=0.2$~s will not be visible. The Fig.~\ref{fig:fig3} for htfr data shows that the sensitivity of the survey varies little and is close to the best possible $S_{min}=4.4$~mJy for pulsars with $P_0>0.5$~s and $DM \le 100$~pc/cm$^3$. From Fig.~\ref{fig:fig3} it can be seen that on $DM > 100$ pc/cm$^3$ the sensitivity when searching for htfr data begins to deteriorate sharply for pulsars with periods $P_0 \ge 0.5$~s, this is due to the scattering effect, which increases the width of the pulse. At the same time, it is known (see figures in papers \citeauthor{Bhat2004} (\citeyear{Bhat2004}), \citeauthor{Kuzmin2007} (\citeyear{Kuzmin2007}), that the amount of scattering on the same $DM$ can differ by three orders of magnitude. In particular, in Fig.~4 of the paper \citeauthor{Bhat2004} (\citeyear{Bhat2004}), at least 4 pulsars are visible at $DM=500$~pc/cm$^3$ for which the scattering is equal to the scattering at $DM=200$~pc/cm$^3$ in the inscribed dependence of the scattering magnitude on the $DM$.  Based on the paper \citeauthor{Bhat2004} (\citeyear{Bhat2004}) and sensitivity evaluations on Fig.~\ref{fig:fig3}, it can be concluded that individual pulsars with narrow pulses (duty cycle is less than $10^{-2}$) and periods of more than 2-3~s can still be observed in the survey at $DM = 1000$~pc/cm$^3$ when the pulse is spread out due to scattering for almost the entire period. Fig.~\ref{fig:fig4} shows a visual representation of the sensitivity difference in the ltfr and htfr data surveys. To obtain the dependencies in the figures, we have taken the ratio of the sensitivities of $S_{min-short}/S_{min-long}$ for one observation session. 

The accumulation of power spectra and periodograms increases sensitivity and shifts the $S_{min}(DM)$ dependencies downwards (Fig.~\ref{fig:fig3}), but does not change the $DM$ boundaries at which it makes sense to search for pulsars. The best sensitivity for pulsars when searching on htfr data is equal to $S_{min-best}=4.4$~mJy$/36=0.12$~mJy in the direction of zenith. In reality, pulsars may not be at the zenith, but their coordinates may fall between the beams of LPA3, and therefore the sensitivity will be worse for almost all directions. If we neglect the dependence of sensitivity on the $DM$ and on the period, then the worst and typical sensitivities will be equal to: $S_{min-worst}=2.5 \times 0.12=0.3$~mJy (when the pulsar is located exactly between the LPA beams); $S_{min-typical}=1.5 \times 0.12=0.18$~mJy.

However, the best sensitivity for the LPA LPI is achieved only for the declination of $55^o$ equal to the latitude of Pushchino, when the observed radio sources are exactly at the zenith. The sensitivity depends on the elevation ($z$) of the source above the antenna when passing the meridian, as $cos(z)$ ($z=\phi-\delta$, where $\phi$ is the latitude of Pushchino, $\delta$ is the declination of observed source). In addition to this, as noted above, the sensitivity deteriorates if the source coordinates fall between the beams of the LPA3 (the coordinates of the LPA3 beams are fixed, see Fig.~1 in the publication \citeauthor{Shishov2016} (\citeyear{Shishov2016})).

\subsection{Comparision PUMPs and other pulsar survey}

For comparison, let's consider the best pulsar search surveys conducted in different frequency ranges with the best sensitivities. In the meter wavelength range, one of the best is the LOTAAS survey conducted on LOFAR. According to \citeauthor{Sanidas2019} (\citeyear{Sanidas2019}), up to a $DM$ of 50~pc/cm$^3$ for pulsars with a period of $0.1-1$~s at a two-hour observation session, the sensitivity is approximately $1.2-1.5$~mJy at a frequency of 135~MHz ($\lambda =2.2$ meters) for observations outside the Galactic plane. At the $DM$ of 100 pc/cm$^3$, the sensitivity, taking into account scattering, deteriorates to $3-5$~mJy. At the $DM$ of 200 pc/cm$^3$, scattering can degrade sensitivity by tens of times. The scattering effect strongly depends on the frequency of observations. Therefore, in the other two surveys, which have a high sensitivity in the search for pulsars, the contributions of scattering to the broadening of pulses up to $DM$ of 200 pc/cm$^3$ can be neglected. In the survey on the 300-meter Arecibo mirror (AO327, \citeauthor{Deneva2013}, \citeyear{Deneva2013}) at a frequency of 327 MHz ($\lambda =0.92$~m), the sensitivity when searching for pulsars with a period of $0.1-1$~s in a minute session is approximately $0.4$~mJy. The survey conducted on a 500-meter FAST mirror (GPPS, \citeauthor{Han2021} (\citeyear{Han2021})) at a frequency of 1250~MHz ($\lambda =0.24$~m) has a very high sensitivity. With the duration of the observation session of five minutes, for pulsars with a period of $0.1-1$~s in the FAST survey, a sensitivity of $0.005-0.01$~mJy was achieved. Assuming that the spectral index in the spectra of all pulsars is $1.7$, it is possible to deduce the sensitivities achieved in pulsar surveys to a frequency of 111~MHz. Let us quote the sensitivity estimates in the best surveys on the search for pulsars having $DM<100$~pc/cm$^3$ and $P>0.5$~s located outside of the plane of the Galaxy: PUMPS – 0.1-0.3~mJy, LOTAAS – 3-5~mJy, AO327 – 2.5~mJy, GPPS – 0.3-0.6~mJy. Fig.~\ref{fig:fig3} shows that the sensitivities of PUMPS and GPPS surveys are comparable and several times higher than the sensitivities achieved in other surveys. 

\subsection{Completeness for low and high time-frequency resolution pulsar survey}

Using the data on Fig.~\ref{fig:fig3} and Fig.~\ref{fig:fig4}, as well as additional calculations (Eq.~\ref{eq:1}), and taking into account the features of LPA3 as an antenna array, it is possible to determine the expected completeness of the survey, that is, the level of flux density at which all pulsars with no variability are guaranteed to be detected in the PUMPS survey. For htfr and ltfr data for the $+21^o < \delta < +42^o$ area, where interference levels are minimal, a completeness is shown in Table~\ref{tab:tab1} for different periods and different DM, assuming that completeness is determined by the worst observation conditions, and the observations themselves take place outside the Galactic plane ($|b|>10^o$). On $DM$ higher than those indicated in columns 4 and 8, the sensitivity (columns 1, 2, 5 and 6) begins to deteriorate sharply. The boundary of the $DM$ was taken at the points where the sensitivity fells 1.5 times from the best in the periods given in columns 3 and 7. We also assumed that a candidate under study has the worst zenith distance ($z=34^o$, for $\delta =+21^o$), and is located exactly between two beams of LPA3 when the received energy is 0.405 of that coming to the antenna. Completeness estimates for observations in the galactic plane will be 2.25 times higher, they are not given in Table~\ref{tab:tab1}.

\begin{table*}
	\centering
	\caption{Completeness levels in the PUMPS survey for ltfr and htfr data when accumulated in one observation session with a duration of $204.8$~s, as well as when accumulated for 4 years ($4 \times 365 \times 204.8$~s for ltfr data) and 7 years (for htfr data).}
	\label{tab:table1}
	\begin{tabular}{ccllccll} 
		\hline
	$S_{min-long}$&$S_{min-long-sum}$&$P$&$DM$&$S_{min-short}$&$S_{min-short-sum}$&$P$& $DM$\\
	(mJy)&(mJy)&(s)&(pc/cm$^{-3}$)&(mJy)&(mJy)&(s)&(pc/cm$^{-3}$)\\
	\hline	
45  &1.25 &0.025 &<20 & -  & -  & - & -\\
30  &0.83 &0.05  &<35 & -  & -  & - & -\\
23  &0.64 &0.1   &<50 & -  & -  & - & -\\
19  &0.53 &0.2   &<70 &51  &1.82&0.2&<30\\
17.5&0.49 &0.3   &<105&36.5&1.30&0.3&<45\\
17  &0.47 &0.5   &<150&26.5&0.95&0.5&<55\\
16.3&0.45 &1.0   &<200&20  &0.71&1.0&<85\\
15.5&0.43 &2.0   &<200&17.5&0.63&2.0&<130\\
13.5&0.38 &3.0   &<200&17  &0.61&3.0&<180\\
12  &0.33 & 5.0  &<200&14  &0.50&5.0&<200\\
		\hline
	\end{tabular}
	\label{tab:tab1}
\end{table*}

If we talk about the efficiency of the search on different $DM$, then the PUMPS survey conducted on ltfr data is maximally effective up to $DM=30-50$~pc/cm$^3$ for periods   $P>0.5$~s. Although, as observations show, pulsars with $DM >200$~pc/cm$^3$ can be detected in it (\citeauthor{Tyulbashev2017}, \citeyear{Tyulbashev2017}). The LOTAAS survey is most effective up to a $DM$ of no higher than 100~pc/cm$^3$, and preferably 50~pc/cm$^3$, for pulsars with $P>0.1$~s. At the same time, we note that a new pulsar PSR J2006+22 was detected in the survey at $DM=130$~pc/cm$^3$, and among the known pulsars, 12 pulsars with $DM>150$~pc/cm$^3$ were found in a blind search, (Table A1 in Appendix, \citeauthor{Sanidas2019} (\citeyear{Sanidas2019})). The surveys AO327 and GPPS are effective for all $DM<200$~pc/cm$^3$, since the sensitivity losses associated with scattering and intra-channel broadening at the frequencies where the surveys were conducted are not significant.

The sensitivity in each of the surveys presented in Fig.~\ref{fig:fig3} exceeds the sensitivity of the surveys made earlier. It would seem that all known pulsars located in the studied areas should be detected in the new surveys. However, this is not the case. In search of pulsars in the LOTAAS survey (see Fig.~8, \citeauthor{Sanidas2019}  (\citeyear{Sanidas2019})) it is claimed that the declared sensitivity and the number of repeated detections are in agreement. The comparison of the number of pulsars in the ATNF catalog ($-3^o < \delta < +75^o$; $DM<200$~pc/cm$^3$; $P_0 > 0.1$~s) and in the table A1 (Appendix A with repeated pulsars detection \citeauthor{Sanidas2019} (\citeyear{Sanidas2019})) shows that LOFAR was able to see about half of the known pulsars. In survey AO327 (\citeauthor{Deneva2013}, \citeyear{Deneva2013}) 44 known pulsars were detected, which is less than 20\% of the known pulsars in the ATNF catalog ($-1^o < \delta < +75^o$; $|b|>5^o$). In the text of the paper on the GPPS survey (\citeauthor{Han2021}, \citeyear{Han2021}) there is no explicit mention of the number of known pulsars that were not detected within the boundaries of the area where the survey was conducted. From the early surveys, note Parks survey  (\citeauthor{Manchester1996}, \citeyear{Manchester1996}). The paper says that 15 out of 161 known pulsars could not be detected in the survey area. That is, 9.3\% of previously detected pulsars were not detected in the survey. Thus, it can be argued that some of the known pulsars, for some reason, are not detected in surveys that are conducted in wider bands and using improved data processing methods. The detection of known pulsars in the PUMPS survey will be discussed in the next paragraph.

The most obvious reasons why pulsars detected in some surveys are not visible in other surveys are: possible cut offs or flattening of spectra at low frequencies, possible refractive and diffractive interstellar scintillations, intrinsic variability of pulsars. For any source visible in one pulsar search survey and not detected in another survey, a special study should be conducted to identify the causes of "invisibility".

The main goal of any new survey is the search for new pulsars. For LOTAAS survey, conducted on LOFAR at the frequency 135 MHz (\citeauthor{vanLeeuwen2010}, (\citeyear{vanLeeuwen2010}), \citeauthor{Stappers2011} (\citeyear{Stappers2011}), \citeauthor{Sanidas2019} (\citeyear{Sanidas2019})), model calculations and estimates of the expected number of new pulsars during the search were made. We have not made a strict estimate of the expected number of new pulsars in PUMPS, but we give here some considerations when comparing the PUMPS and LOTAAS surveys:

a) Close latitudes of the LOFAR core location ($\phi =53^o$) and LPA LPI antennas ($\phi =55^o$) lead to very close dependencies of the sensitivity reduction associated with the zenith angle.

b) PUMPS sensitivity, when summing spectra and periodograms in long data accumulated over an interval of more than 7 years, significantly exceeds LOTAAS sensitivity at any available $DM$ and at periods $P_0 \ge 25$~ms (see Fig.~\ref{fig:fig3}).

c) A number of pulsars in ATNF, having the period $P_0 < 25$~ms and theoretically available for detection on LOTAAS, but not available for detection in PUMPS, it is relatively small - it makes 13\% of the total number of all ATNF pulsars.

d) Survey areas of PUMPS and LOTAAS are comparable - this is 17000 sq.deg. and 20000 sq.deg (the entire northern hemisphere).

Based on this, we can conclude that the PUMPS survey should be better than the LOTAAS survey for the task of blind pulsar search. According to a special \citeauthor{vanLeeuwen2010} (\citeyear{vanLeeuwen2010}) study, 900 new pulsars are expected to be detected in the LOFAR survey. Considering that the PUMPS survey is  superior to LOTAAS on almost all parameters related to the minimally detectable pulsar flux densities for different pulsar periods and different $DM$, we can expect no fewer new pulsars in our survey.

\section{Discussion}

In the paragraph considering the sensitivity of PUMPS, it was said that several tens of percent of known pulsars falling into the survey area are usually recorded in the surveys area conducted. For example, in the survey conducted on a 64-meter telescope in Parks (\citeauthor{Manchester1996}, \citeyear{Manchester1996}), in a blind search, more than 90\% of all pulsars in the area were detected, and this survey turned out to be the best in re-detecting known pulsars. The sensitivity of our ltfr data search is high (see Fig.~\ref{fig:fig3}), and we can expect that all known pulsars will be detected in it.

As is known, pulsars have steep spectra, but if we talk about observations in the meter range, then flattening of the spectra is often observed, or even cut-off (\citeauthor{Malofeev1996}, \citeyear{Malofeev1996}, \citeauthor{Bilous2016}, \citeyear{Bilous2016}). In the case of a spectrum cut-off, a pulsar visible at high frequencies may not be observed at low frequencies. Therefore, it is more correct to compare the number of observed pulsars in surveys that were conducted at close frequencies. In this case, it can be expected that even with an erroneous spectral index, the recalculation of the flux density will not lead to large errors. LOTAAS survey (\citeauthor{Sanidas2019}, \citeyear{Sanidas2019}), conducted on LOFAR is close by the central frequency of observations to the PUMPS survey conducted at LPA3. According to Fig.~\ref{fig:fig3}, the sensitivity of PUMPS is better than the sensitivity of LOTAAS, and this allows us to hope that all pulsars that have ever been observed in the meter wavelength range will be detected in our observations.

We have chosen for testing the area $+21^o < \delta < +42^o$, where there is a minimum of interference observed, and the area itself is as close to the zenith as possible. In Fig.~\ref{fig:fig3}, the sensitivity in the PUMPS survey according to htfr data is from several times to an order of magnitude better than the sensitivity in the LOTAAS survey. The shorter the pulsar period, and the greater its $DM$, the smaller is the gain in sensitivity. For verification, all pulsars discovered in LOTAAS survey that fall into the area were taken for consideration, regardless of their period and $DM$. A total of 33 pulsars from the LOTAAS survey fall into the area. These pulsars have a $P_0$ from 33 ms (PSR J1658+36) up to 2.42 s (PSR J0349+23) and $DM$ from 3.05~pc/cm$^3$ (PSR J1658+36) up to 130.56~pc/cm$^3$ (PSR J2006+22). For a part of the pulsars, harmonics in the summed-up power spectra obtained from ltfr data, were not detected. For all LOTAAS pulsars, we have made a blind search for htfr data, based on the known coordinates, and trying to detect the pulsar in the same way as it is done for any pulsars when searching. 

\begin{figure}
	\includegraphics[width=\columnwidth]{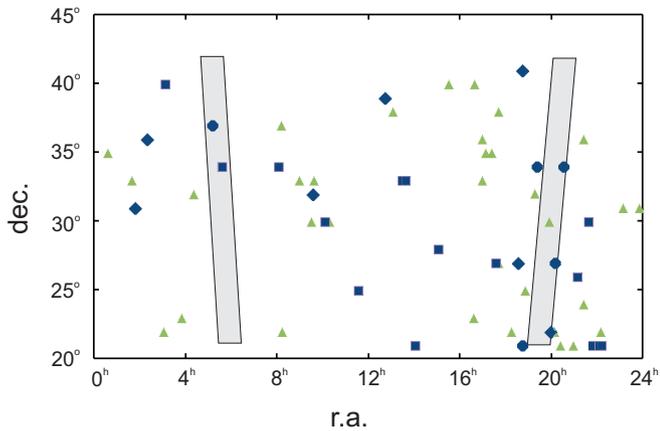}
    \caption{Green triangles are new pulsars detected both in the PUMPS survey and in the LOTAAS survey. The blue circles are the new pulsars found in PUMPS (Table~2). The blue diamonds and squares are pulsars and rotating transients detected in the PUMPS survey earlier. The declination and right ascension of pulsars are shown on the vertical and horizontal axes. The gray color in the figure schematically shows the plane of the Galaxy within the limits $|b|<10^o$.}
    \label{fig:fig7}
\end{figure}

Fig.~\ref{fig:fig7} shows with marks of different colors pulsars from PUMPS and LOTAAS surveys falling into the area. In total, the figure shows the location of 60 pulsars. PUMPS was detected 100\% new LOTAAS pulsars in the area, regardless of their $P_0$ and $DM$. There are 11 new pulsars in the same area, published in this paper and in total, Fig.~\ref{fig:fig7} shows 20 new pulsars and transients discovered in PUMPS (see our website\footnote{https://bsa-analytics.prao.ru/en/}), but not detected in the LOFAR survey. Almost all open pulsars are located outside the galactic plane. Apparently, the new search for pulsars in the htfr data will be most effective when searching outside the plane of the Galaxy.

Additional verification of PUMPS real sensitivity was done on ATNF pulsars, included into the area $+21^o <\delta < +42^o$, having $P_0>0.5$~s and $DM<100$~pc/cm$^3$. In ATNF catalogue for the end of 2020, there are 100 pulsars the listed in this area (see Appendix~B). According to ATNF, 7 sources are RRATs (see Appendix~C), and their regular (pulsar) emission has not yet been detected.  These RRATs are excluded from the list which determines the proportion of detected pulsars. Out of the remaining 93 pulsars, for 13 pulsars in the summed-up power spectra obtained from ltfr data, periodic emission was not detected. 11 pulsars (see Appendix~C) out of these 13 were not detected in LOTAAS also. Seven pulsars out of 13 have ATNF estimates of the flux density at 400~MHz. The expected flux densities at 111~MHz when recalculated with a spectral index of 1.7 show values from 3.5 to 53~mJy with a median flux density of about 18~mJy. That is, almost all sources should be detected on LPA3/LOFAR in single observation sessions.

If we assume that 13 pulsars out of 93 cannot be detected, then the proportion of detections in a blind search is defined as 86\%. It seems that, it is possible to exclude from these 13 pulsars 11 pulsars that were not detected either on the LPA LPI or on LOFAR, assuming that for some objective reasons these pulsars are not observed in the meter wavelength range. Then there are only 2 pulsars remaining (PSR J1919+2621; PSR J2030+2228) out of 93, which cannot be detected on the LPA. In this case, the proportion of ATNF pulsars detected by blind search in PUMPS increases to 97.8\%. To find out the reasons why for 11 pulsars periodic emission in the meter wavelength range is not detected, a special study is required.  First of all, we need multi-frequency simultaneous observations for these pulsars.

40 out of 100 pulsars of Appendix~B were discovered in a blind search by different authors for the period from 2014 to the present. Taking into account the present paper and the search works published by us during the same period, the number of new pulsars detected in PUMPS has increased to 42. Thus, the total number of pulsars located in the area ($+21^o < \delta < +42^o$; $P_0>0.5$~s; $DM<100$~pc/cm$^3$) has almost doubled in the last 7 years. 

In the section "Observations" it was noted that earlier the search for pulsars was carried out both for classical pulsars with regular (periodic) emission, and for transients, for which the search for individual dispersed pulses was done. At the website\footnote{https://bsa-analytics.prao.ru/en/} many known pulsars were simultaneously in different tables, as they were detected using different search methods. Combining all the tables of the website, we get that as of September 2021, 165 known pulsars were detected in the PUMPS blind search, 73 of them were also detected by individual pulses, without taking into account the detection of known pulsars in the side lobes of the LPA LPI. 29 known pulsars were detected only by individual pulses. A special search for their periodic emission was not carried out. The survey also revealed 42 new pulsars, of which 6 were also detected by individual pulses. 46 new RRAT pulsars were also detected. 

There is the obvious problem with the expected and detected number of new pulsars. The LOTAAS survey revealed less than 10\% of the expected number of new pulsars. A similar percentage of discoveries is observed in PUMPS. Thus, the two low-frequency surveys together revealed about one sixth of the pulsars out of the expected 900 (\citeauthor{vanLeeuwen2010}, \citeyear{vanLeeuwen2010}). Such a strong discrepancy between the expected and actually discovered number of pulsars may indicate the complete exhaustion of pulsars with $DM \le 100$~pc/cm$^3$. This apparent discrepancy is not the subject of this paper. We hope to answer the question about the possible exhaustion of pulsars located at distances from the Earth not exceeding several kiloparsecs, after searching for pulsars from htfr data, in which sensitivity will be realized an order of magnitude better than in the LOTAAS survey.

\section{Conclusion}

In this paper, a number of results is obtained:

1) We present the discovery of 11 new pulsars. For 10 pulsars, we were not able to create average profiles. Taking into account previous papers, 42 new pulsars with regular emission have already been discovered in the survey, and in total, 195 known pulsars and 88 new pulsars and RRATs were blindly detected/redetected in the PUMPS survey when processing data obtained for 2014-2018 years. The total number of new pulsars detected in PUMPS at low and high time-frequency resolution is comparable to the number of pulsars detected in LOFAR, which indicates the high efficiency of the PUMPS survey. 

2) The expected (theoretical) sensitivity of the $S_{min-sum-long}$ PUMPS survey with the summation of power spectra and periodograms using 32-channel data and a sampling time of 12.5 ms per raw data point is approximately an order of magnitude better than the sensitivity of the $S_{min}$ of all previous and current surveys, with the exception of the GPPS survey (\citeauthor{Han2021}, \citeyear{Han2021}) conducted on a 500-meter FAST telescope. The best theoretical sensitivity in the PUMPS survey is 3-4 times better than in the GPPS survey for pulsars with $P_0 \ge 0.5$~s and $DM \le 100$~pc/cm$^3$. At the same time, the total signal accumulation time in PUMPS is equivalent to six days of continuous observations at a frequency 111 MHz. The sensitivity declared in the PUMPS survey is expected if the pulsar lacks any variability, and the interference is removed so well that in practice they do not introduce degradation into the final sensitivity;

3)	Experimental verification shows that PUMPS objectively has a higher sensitivity than LOTAAS; 

4)	The completeness of the planned survey differs for different $DM$ and periods. It can be expected that the PUMPS survey will detect all pulsars that have $P_0 \ge 0.5$~s and $DM \le 100$~pc/cm$^3$, located outside the plane of the Galaxy in the area $+21^o < \delta < +42^o$, and having a flux density $S>0.5$ mJy at the frequency 111 MHz;

5)	A large number of known ATNF pulsars, not detected either in LOFAR observations or in LPA3 observations, indicates the objective absence (or extreme weakness) of signals from some of these pulsars in the meter wavelength range in the time interval 2014-2019 years;

6) The blind survey with the high time-frequency resolution data is underway.

\section*{Data availability}
The PUMPS survey not finished yet. The raw data underlying this paper will be shared on reasonable request to the corresponding author. The additional figures of detected pulsars in http://prao.ru/online\%20data/onlinedata.html

\section*{Acknowledgements}

The authors are grateful to the antenna group for their active assistance in conducting observations; to L.B. Potapova for help with the preparation of some figures; to V.S. Tyul’bashev and Yu.A. Belyatsky for maintaining the server in working condition and constant consultations; M.S. Burgin for provided server for processing of observations.




\newpage
\appendix

\section{Data analysis on the example of J1745+12 pulsar}

\begin{figure*}
	\includegraphics[width=\textwidth]{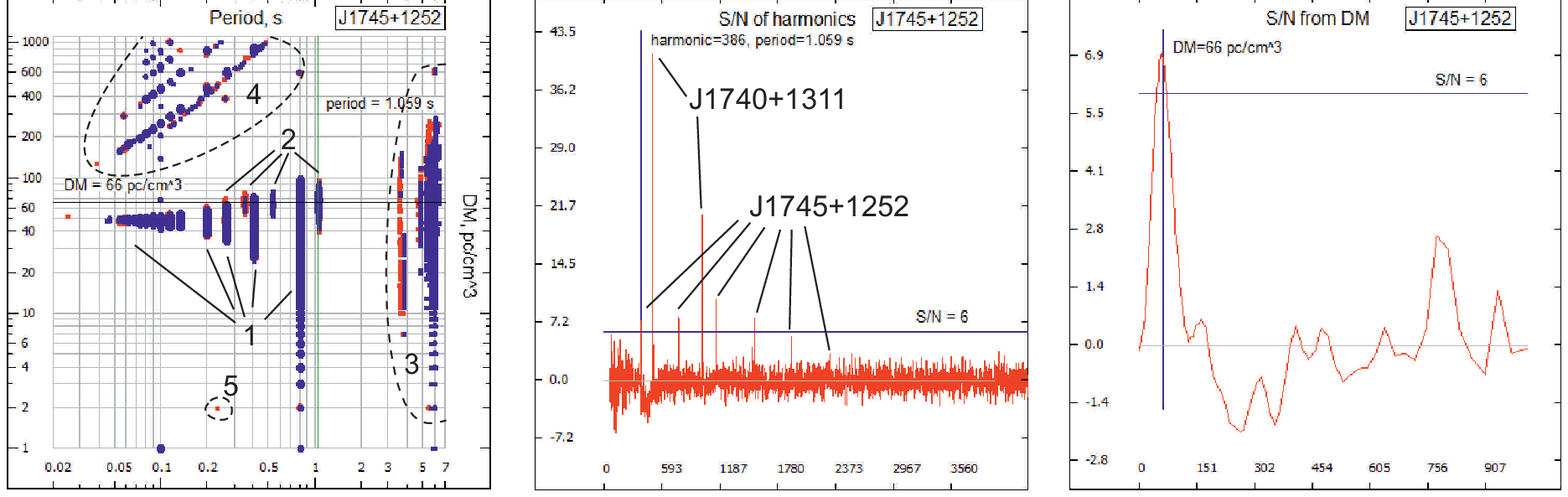}
    \caption{Pictures generated in the central part of the working window of the program. }
    \label{fig:fig2}
\end{figure*}

Fig.~\ref{fig:fig2} shows from left to right the diagrams generated by the program: a map with visualization of the power spectra summed-up over 5.5 years, the power spectrum of our point of interest ($P_0, DM$) on the map, and a diagram of the dependence of the harmonic height $P_0$ vs. the sorted-out $DM$. For the demonstration, we have chosen the pulsar J1745+12, published in the papers \citeauthor{Sanidas2019} (\citeyear{Sanidas2019}), \citeauthor{Tyulbashev2020} (\citeyear{Tyulbashev2020}). This pulsar was one of the candidates detected from short data in the power spectra summed-up over a four-year interval. Previously, we were not able to find a single recording in the htfr data to give a $DM$ estimate for it and to construct an average profile. 

The map (left part of Fig.~\ref{fig:fig2}) shows the visualization of power spectra obtained from htfr data, and summed-up for about 800 days. Circles of different diameters show the harmonics of these spectra with $S/N \ge 6$. There are 5 characteristic areas on the map, two of which look like sets of vertical strips characteristic for pulsars. 

The area~1 consists of a set of 15 vertical strips. The dispersion   $DM=49\pm 1$ pc/cm$^3$ and period $P_0=0.804$~s are determined from the figure. It turned out that on a right ascension $\alpha_{2000}=17^h45^m$, corresponding to the time of passage through the meridian, at the edge of the LPA LPI diagram, remnants of radiation from a known pulsar PSR J1740+1311  are observed for which according to ATNF, $DM=48.668$ pc/cm$^3$, and $P_0=0.80305$~s.

The area~2 consists of 4 vertical strips. Pulsar characteristics determined using the visualization program: $\alpha_{2000}=17^h45^m \pm 1.5^m$, $\delta_{2000}=13^o02^\prime \pm 20^\prime$, $P_0=1.059 \pm 0.001$~s, $DM=66.5 \pm 1$~pc/cm$^3$. As we mentioned above, this pulsar candidate was detected in our observations, did not have a $DM$ estimate, and was published in the paper \citet{Sanidas2019}. According to this paper, the pulsar characteristics PSR J1745+12 are $\alpha_{2000}=17^h45^m44^s$, $\delta_{2000}=12^o52^\prime 38^{\prime \prime}$, $P_0=1.0599$~s, $DM=66.141$~pc/cm$^3$.

The other highlighted areas in the figure indicate: 3 - low-frequency noise limiting the maximum periods of the map; 4 - diagonal lines generated by pulsars and periodic interference; 5 - an example of periodic interference observed at $S/N \ge 6$. Horizontal and vertical lines are estimated period and $DM$ of pulsar PSR J1745+1252.

Estimates of pulsar flux density at frequencies 128-1532~MHz were given in the paper \citeauthor{Tan2020} (\citeyear{Tan2020}): $S_{128}=9.6$~mJy; $S_{167}=5.5$~mJy; $S_{334}=2.2$~mJy; $S_{1532} < 0.08$~mJy. Using the spectral index given in the paper $\alpha = 1.5$ ($S \sim \nu^{-\alpha}$), it is possible to estimate the expected flux density at the central frequency of observations of LPA3 ($S_{110.3}=12$~mJy). The source is located in the extragalactic plane (galactic latitude is $b=+20.3^o$). Sensitivity estimation for seconds duration pulsars located outside the Galactic plane is $S_{min}=6-8$~mJy (\citeauthor{Tyulbashev2016}, \citeyear{Tyulbashev2016}). Therefore, PSR J1745+1252 must be visible in single observation sessions. However, considering that the source is not observed at the zenith, and the coordinates of the beam in declination do not coincide with the declination of the source, we estimated the expected sensitivity of LPA3 in the direction of the source as 12-16~mJy. Thus, this source is at the sensitivity limit of the LPA3 radio telescope when observed in one session in the direction of the source. In the summed-up power spectra obtained from the raw data, the sensitivity has increased by about 25 times, and therefore the pulsar PSR J1745+1252 is clearly visible on the map. The map shows signals with $S/N \ge 6$. Initially, signals with $S/N \ge 4$ are applied onto the map, therefore, there are a lot of "garbage" points on such maps, indicating the observed weak periodic signals (see Fig.~\ref{fig:fig1}). But we can increase the lower limit of the $S/N$ harmonics shown on the map to cut off the points we don't need and make the picture more compelling. In area~5, we see an example of a single point with interference remaining after clipping. In area 3, low-frequency noises are visible for periods longer than 3 seconds, in area 4 there are diagonal lines generated by both pulsars visible on the map and interference. 

In the summed-up power spectrum (middle part of Fig.~\ref{fig:fig2}) obtained for raw data collected at  $DM=66$~pc/cm$^3$, two sets of harmonics associated with pulsars PSR J1740+1311 and PSR J1745+1252 are visible. For pulsar PSR J1740+1311, only two of the fifteen harmonics visible on the map are observed in the power spectrum. This is due to the fact that the $DM$ of this pulsar is markedly different from the $DM$ at which the power spectrum is obtained. In the second set, six harmonics are visible, four of which have $S/N \ge 6$. On the original map showing signals with $S/N \ge 4$, and in the presented picture of the power spectrum, the fifth harmonic is visible ($S/N=5.3$). A weak sixth harmonic is visible in the power spectrum, but its $S/N$ is approximately 3, and therefore it is not visible on the map. 

At the beginning of the power spectrum, there is a broadening of the noise track and poor subtraction of the baseline. We were unable to find a dependence suitable for the automatic correct subtraction of low-frequency noise in the summed-up spectra. This is due to the fact that in the first few hundred points of the summed-up power spectra, a nonlinear dependence is observed. Practical processing shows that the search for pulsars in our data using power spectra works well for periods not exceeding 3-4 seconds. Fragment of the summed-up power spectrum length about 4500 points out of 16384 points is in middle part of~Fig.~\ref{fig:fig2}. On the vertical axis is the harmonic amplitude in units of $S/N$, on the horizontal axis is the number of the point in the power spectrum. The vertical blue line continuing the first harmonic shows the current value of $P_0$, (it was this harmonic that created the selected point on the map). We do not sum up the amplitudes of harmonics in the program, but when checking candidates, we pay attention to harmonics having $S/N \ge 4$. The additional harmonics visible at $S/N < 6$ and the limited area of the $DM$ region on which they appear are for us an indirect sign of real pulsars. The horizontal blue line in the figure marks $S/N=6$.

The right part of Fig.~\ref{fig:fig2} shows the dependence of the height of the first harmonic of the pulsar PSR J1745+12 in $S/N$ units on the $DM$ being checked. This diagram shows that the maximum of the first harmonic of PSR J1745+12 rises to $S/N=6.9$ to $DM = 66$~pc/cm$^3$, and then decreases. The horizontal blue line marks $S/N=6$.

\section{Pulsars detected in the side lobes of LPA3 in the PUMPS survey}

The list is presented pulsars detected in the side lobes of LPA3 in the PUMPS survey. The pulsar not previously detected in the side lobes is marked in bold. For some of the pulsars, there are many side lobes observed located at distances up to tens of degrees in right ascension and declination away from the coordinates of the pulsar that give rise to these lobes. According to ATNF, only 2 pulsars out of 19 have a flux density $S_{400} <15$~mJy at the frequency 400 MHz. Median value of the flux density of the sources forming the side lobes is $S_{med-400}=73$~mJy (PSR J0826+2637). If we assume a spectral index of 1.7, then when converted to 111~MHz, the expected flux density of PSR J0826+2637 will be $S_{111}=645$~mJy. That is, very strong pulsars are mainly observed in the side lobes.

PSR J0048+3412; PSR J0137+1654; PSR J0323+3944; PSR J0332+5434; PSR J0814+7429; PSR J0826+2637; PSR J0922+0638; PSR J0953+0755; PSR J1136+1551; PSR J1239+2453; PSR J1509+5531; PSR J1543+0929; PSR J1645-0317; PSR J1735+0724; PSR J1740+1311; PSR J1921+2153; PSR J2018+28; PSR J2022+2854; PSR J2305+3100.

\section{Known pulsars in the studied area}

The list of known pulsars ($+21^o <\delta < +42^o$; $P_0 >0.5$~s; $DM < 100$~pc/cm$^3$): a) previously detected in a blind search as ordinary seconds duration pulsars (\citeauthor{Tyulbashev2016}, \citeyear{Tyulbashev2016}, \citeauthor{Tyulbashev2017}, \citeyear{Tyulbashev2017}, \citeauthor{Tyulbashev2020}, \citeyear{Tyulbashev2020})); they are indicated by a superscript digit 1; b) detected for the first time at the frequency 111~MHz using summed power spectra or summed periodograms (this paper), and also discovered as rotating transients (\citeauthor{Tyulbashev2018a}, \citeyear{Tyulbashev2018a}, \citeauthor{Tyulbashev2018b}, \citeyear{Tyulbashev2018b}), having also ordinary (periodic) pulsar emission are indicated by a superscript digit 2; c) caught in the area, but not detected in the search by short data. Detected by blind search by htfr data in individual power spectra or periodograms (indicated by a superscript digit 3). 

Known pulsars that could not be detected were left without superscript designations. The bold font shows pulsars, the articles on the detection of which appeared after 2013, i.e. after the start of the search for pulsars on the LPA LPI. The pulsars and RRATs discovered by the  LPA LPI and published in the period 2016-2020 are  indicated  in bold italics. For all pulsars found on the site\footnote{https://bsa-analytics.prao.ru/en/pulsars/known.php} drawings confirming their detection are posted. Out of the 100 pulsars presented below, 40 pulsars appeared in publications after 2013.

RRAT without pulsar emission:

\textit{ \textbf{\textbf{PSR} J0139+3336; \textbf{PSR} J0302+2252$^1$; \textbf{PSR} J1132+25; \textbf{PSR} J1336+33; \textbf{PSR} J1502+28; \textbf{PSR} J2209+22$^2$;}} PSR J2225+35

Canonical second pulsars:

\textbf{\textbf{PSR} J0039+35$^2$}; PSR J0048+3412$^1$; \textit{ \textbf{\textbf{PSR} J0146+31$^1$}}; PSR J0156+3949$^2$; PSR J0158+21$^1$; \textit{ \textbf{ \textbf{PSR} J0220+36$^1$}}; PSR J0323+3944$^1$; \textbf{\textbf{PSR} J0349+2340$^2$}; PSR J0417+35$^1$; \textit{ \textbf{\textbf{PSR} J0421+3255$^1$}}; PSR J0457+23$^1$; PSR J0528+2200$^1$; PSR J0540+3207$^1$;  PSR J0546+2441$^1$; PSR J0555+3948$^3$; PSR J0611+30$^1$; \textbf{\textbf{PSR} J0613+3731$^1$}; PSR J0754+3231$^1$; \textit{ \textbf{\textbf{PSR} J0811+37$^1$}}; \textbf{\textbf{PSR} J0813+22$^2$}; PSR J0826+2637$^1$; PSR J0927+2345; \textit{ \textbf{\textbf{PSR} J0928+30$^1$}}; \textit{ \textbf{\textbf{PSR} J0935+33$^1$}}; PSR J0943+2253$^1$; \textbf{\textbf{PSR} J0944+4106$^1$}; PSR J0947+2740$^1$; PSR J1238+2152$^1$; PSR J1239+2453$^1$; \textit{ \textbf{\textbf{PSR} J1242+39$^1$}}; PSR J1503+2111; PSR J1532+2745$^1$; \textbf{\textbf{PSR} J1538+2345$^1$}; PSR J1549+2113$^2$; \textit{ \textbf{\textbf{PSR} J1635+2332$^1$}}; \textit{ \textbf{\textbf{PSR} J1638+4005$^1$}}; PSR J1649+2533$^1$; PSR J1652+2651$^1$; \textit{ \textbf{\textbf{PSR} J1657+3304$^1$}}; PSR J1720+2150$^2$; \textit{ \textbf{\textbf{PSR} J1722+35$^1$}}; \textbf{\textbf{PSR} J1740+27$^2$}; PSR J1741+2758$^1$; \textbf{\textbf{PSR} J1741+3855$^2$}; PSR J1746+2245$^3$; PSR J1746+2540$^1$; PSR J1758+3030$^1$; PSR J1813+4013$^1$; \textbf{\textbf{PSR} J1821+4147$^1$}; \textit{ \textbf{\textbf{PSR} J1832+27$^1$}}; \textit{ \textbf{\textbf{PSR} J1844+41$^1$}}; \textbf{\textbf{PSR} J1849+2559$^2$}; PSR J1900+30$^2$; PSR J1907+4002$^1$; PSR J1912+2104; PSR J1912+2525$^1$; PSR J1913+3732$^1$; \textbf{\textbf{PSR} J1916+3224}; \textbf{\textbf{PSR} J1919+2621}; PSR J1920+2650$^1$; PSR J1921+2153$^1$; PSR J1929+2121; \textbf{\textbf{PSR} J1929+3817$^1$}; PSR J1931+30; PSR J1946+24; \textit{ \textbf{\textbf{PSR} J1953+30$^1$}}; PSR J2008+2513$^2$; PSR J2015+2524; PSR J2018+2839$^1$; \textbf{\textbf{PSR} J2022+21$^2$}; PSR J2030+2228; PSR J2036+2835$^2$; PSR J2037+3621; \textbf{\textbf{PSR} J2044+28}; PSR J2055+2209$^1$; \textbf{\textbf{PSR} J2057+21$^2$}; PSR J2102+38$^1$; PSR J2111+2106; PSR J2113+2754$^1$; \textbf{\textbf{PSR} J2122+2426$^2$}; PSR J2139+2242$^1$; PSR J2151+2315$^1$; PSR J2155+2813$^1$; PSR J2157+4017$^1$; \textbf{\textbf{PSR} J2208+4056$^1$}; PSR J2212+2933$^1$; PSR J2227+30$^1$; PSR J2234+2114$^1$; PSR J2305+3100$^1$; PSR J2307+2225; PSR J2317+2149$^1$; \textit{ \textbf{\textbf{PSR} J2350+31$^1$}}; \textbf{\textbf{PSR} J2355+2246$^2$} 

\section{Pulsars marked in the Fig.7}

Pulsars detected in PUMPS and not detected in LOTAAS:

PSR J0146+31, PSR J0220+36, PSR J0509+37, PSR J0933+32, PSR J1242+39, PSR J1832+27, PSR J1843+21, PSR J1844+41, PSR J1921+34, PSR J1958+22, PSR J2008+27, PSR J2030+34

Pulsars detected in both surveys: 

PSR J0039+35, PSR J0139+33, PSR J0302+22, PSR J0349+23, PSR J0421+32, PSR J0811+37, PSR J0813+22, PSR J0857+33, PSR J0928+30, PSR J0935+33, PSR J1303+38, PSR J1529+40, PSR J1635+23, PSR J1638+40,PSR J1657+33, PSR J1658+36, PSR J1707+35, PSR J1722+35, PSR J1741+38, PSR J1740+27, PSR J1916+32 PSR J1814+22, PSR J1849+25, PSR J1953+30, PSR J2006+22, PSR J2022+21, PSR J2057+21, PSR J2122+24, PSR J2123+36, PSR J2209+22, PSR J2306+31, PSR J2350+31

Other marks in Fig.~\ref{fig:fig6} are RRATs from \citep{Tyulbashev2018a, Tyulbashev2018b} (\citeauthor{Tyulbashev2018a}, \citeyear{Tyulbashev2018a}).


\bsp	
\label{lastpage}
\end{document}